# A Radio-fingerprinting-based Vehicle Classification System for Intelligent Traffic Control in Smart Cities

**Benjamin Sliwa[1], Marcus Haferkamp[1], Manar Al-Askary[1], Dennis Dorn[2] and Christian Wietfeld[1]**
[1]Communication Networks Institute, TU Dortmund University, 44227 Dortmund, Germany
e-mail: {Benjamin.Sliwa, Marcus.Haferkamp, Manar.Askary, Christian.Wietfeld}@tu-dortmund.de
[2]Wilhelm Schröder GmbH, 58849 Herscheid, Germany
e-mail: {Dorn}@mfds.eu

*Abstract*—**The measurement and provision of precise and up-to-date traffic-related key performance indicators is a key element and crucial factor for intelligent traffic control systems in upcoming smart cities. The street network is considered as a highly-dynamic Cyber Physical System (CPS) where measured information forms the foundation for dynamic control methods aiming to optimize the overall system state. Apart from global system parameters like traffic flow and density, specific data, such as velocity of individual vehicles as well as vehicle type information, can be leveraged for highly sophisticated traffic control methods like dynamic type-specific lane assignments. Consequently, solutions for acquiring these kinds of information are required and have to comply with strict requirements ranging from accuracy over cost-efficiency to privacy preservation. In this paper, we present a system for classifying vehicles based on their radio-fingerprint. In contrast to other approaches, the proposed system is able to provide real-time capable and precise vehicle classification as well as cost-efficient installation and maintenance, privacy preservation and weather independence. The system performance in terms of accuracy and resource-efficiency is evaluated in the field using comprehensive measurements. Using a machine learning based approach, the resulting success ratio for classifying cars and trucks is above 99%.**

## I. Introduction and Related Work

Intelligent traffic control systems will be a key component of future smart city Intelligent Transportation Systems (ITS) [1] that use means for dynamic direction and type-specific lane assignments in order to cope with an enormous traffic demand. In this context, the street network is interpreted as a CPS where measured information (e.g. traffic flow and density, jam occurrence, etc.) becomes the key for optimizing the overall system state. Consequently, accurate and real-time capable detection and classification systems are required in order to guarantee the efficiency of the overall traffic control system. In this paper, we present a vehicle classification system based on radio-fingerprinting. While our previous work focused on specific aspects like detection of wrong-way drivers within the Wireless Detection and Warning System (WDWS) [2] and in-depth analysis of the classification performance of different machine learning models [3], we provide a system-level perspective and an overall vision in this paper. Furthermore, the resource-efficiency is evaluated as the overall goal to analyze big data on small devices. A system-level overview for the proposed classification system as a source for accurate traffic data is shown in Fig. 1.

### A. Use-cases and Challenges

In addition to the tasks of existing traffic monitoring systems, such as the evaluation of traffic flow and density, future smart city traffic management systems will provide concrete data on individual vehicles (e.g. type, shape and speed). This information is also valuable for systems for smart toll collection and parking space accounting. The detailed evaluation of the necessary information requires highly accurate and fine-grained measuring systems which must simultaneously comply with the following conditions:

1) **Privacy-preservation:** In many countries, the protection of privacy-related data is of enormous importance and is therefore strictly regulated by the legislator. Consequently, sensitive data may only be analyzed and stored temporarily under strict conditions within the intended application. In particular, this applies to applications in public road traffic, where privacy protection for a large number of road users must be ensured.
2) **Cost-efficiency:** Since the smart city context is a large-scale deployment scenario, the cost for installation and

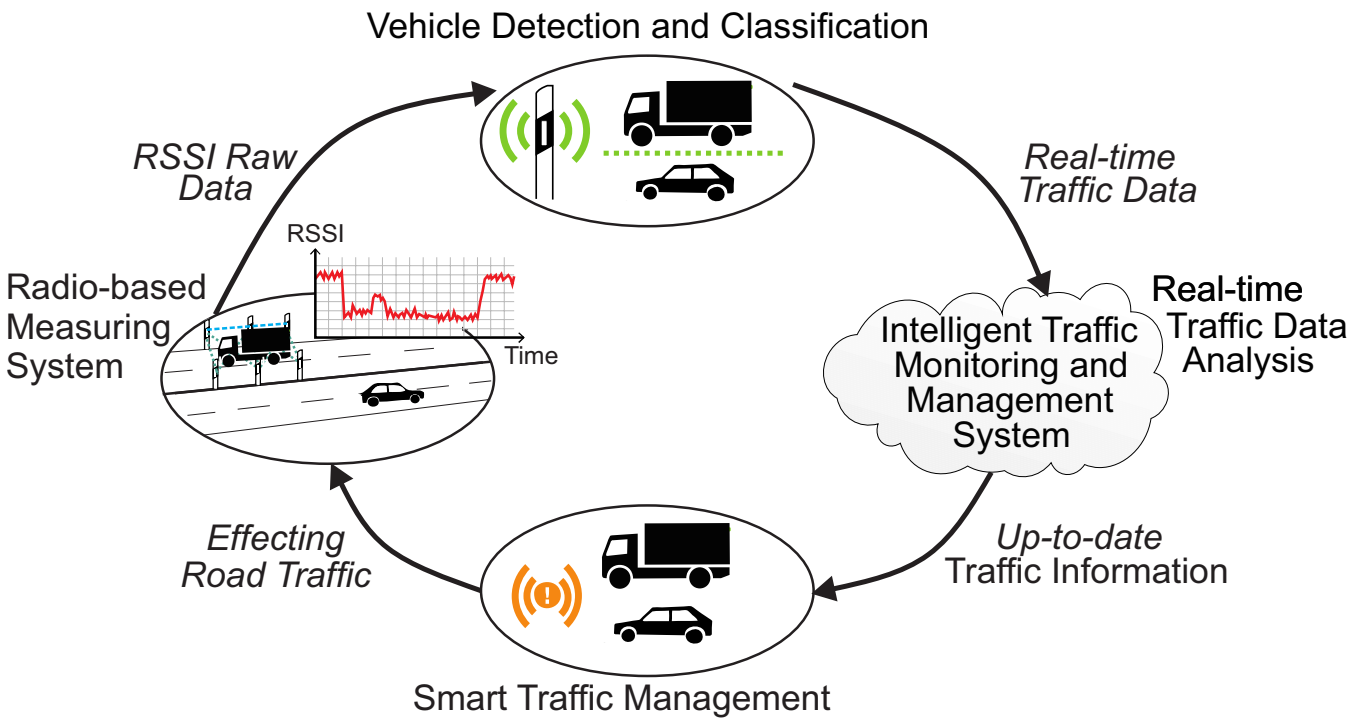

Fig. 1. Application scenario for the proposed classification system as a real-time traffic data source for ITS in a smart city context.

maintenance of the system is a crucial factor. In order to obtain accurate data for high-precision traffic information and forecasts, the sensor network must be installed as widely as possible with a focus on neuralgic traffic junctions. In this context, *invasive* and *non-invasive* solutions can be distinguished depending on if they need to be installed within the road itself (e.g. induction loops) or not. Additionally, the required power supply methods differ from systems connected to the power grid, over battery-based approaches to fully self-sufficient systems.

3) **Weather-independency:** A reliable system must deliver accurate measurement results at all times of the year, despite diverse weather conditions or installation locations. For example, the use of camera-based systems requires a clear view, which cannot be guaranteed at all times (e.g. due to fog, dirt, darkness).

### *B. Related Work*

Various technological solutions are used for gathering classification data directly on the street [4]. For example, camera-based systems using multiple visual units with different perspectives on the road are widely used for vehicle detection and classification. In [5] an enhanced, monolithic visual system using only one camera for vehicle classification is presented. In addition to camera-based systems, methods using laser scanners [6], acoustic sensors [7], accelerometers [8], magnetometers [9] and even Global Positioning System (GPS)-based approaches [10] are further existing solution approaches. Nevertheless, none of the systems is able to completely comply with the above mentioned requirements. This includes, for example, a high susceptibility to errors caused by bad weather conditions in case of heavy rain or dense fog and privacy-related issues for camera-based systems. Furthermore, many of the discussed approaches are invasive and require heavy road construction work for installation or maintenance purposes (e.g. pavement cut) with the effect of extremely high costs for massive deployment scenarios. Considering all mentioned requirements, the proposed radio-based approach provides a favorable tradeoff because it does not violate privacy-constrains, it is cost-efficient to install in a non-invasive way through integration into existing delineator posts and its performance is not significantly influenced by weather conditions. In addition, our proposed solution can also be used for several application scenarios including the detection of wrong-way drivers on motorways as well as traffic and parking space balancing.

The structure of this paper is as follows. After the related work has been discussed, we present the setup of our radio-based solution proposal for vehicle detection and classification in Sec. II. Sec. III provides an overview of the first results achieved with our proposed system as well as evaluations of the resource-efficiency. Finally, the main conclusions of the project are summarized in Sec. IV.

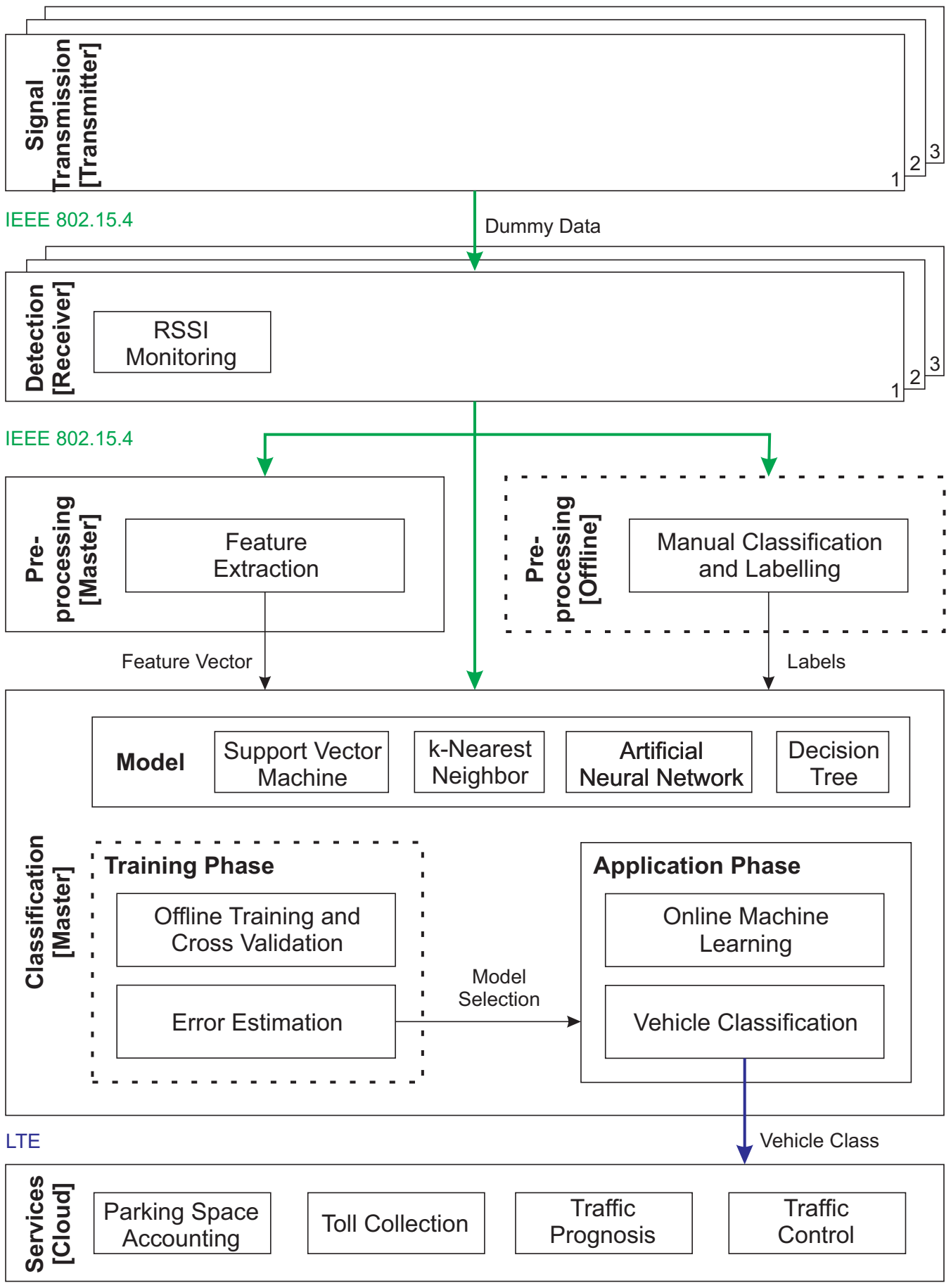

Fig. 2. Architecture model of the proposed system. The dashed components are only required during the offline training phase of the classification models.

## II. Solution Approach and System Architecture

The overall architecture model for the proposed system is illustrated in Fig. 2. In the subsequent subsections, the different subsystems are described with respect to their function for the overall system. Fig. 3 shows the setup used for the field-test evaluations and illustrates the different radio-links $\Phi_i$ of the system. Furthermore, an example trace of a vehicle passing the setup is shown. The different attenuation phases form the *radio-fingerprint* of the vehicle based on its shape and can be regarded as characteristic for the vehicle class (car or truck).

### *A. Transmitter and Receiver Posts*

The delineator posts are the key components of the proposed system. Three posts are mounted on one side of the street and act as *transmitters*, while three other posts being used as *receivers* are mounted on the opposite street side. Within each post, an embedded computer equipped with an IEEE 802.15.4 low power radio interface is installed. Using a token-based channel access method, the transmitters are used as wireless beacons that transmit dummy data. The receiver posts receive the signal from the transmitters and continuously measure the resulting Received Signal Strength Indicator (RSSI) level. For each signal $\Phi_i$, the RSSI sequence is monitored and forwarded as raw data to the master gateway using a IEEE 802.15.4 link. Tab. I shows the most relevant system parameters for the delineator post installation.

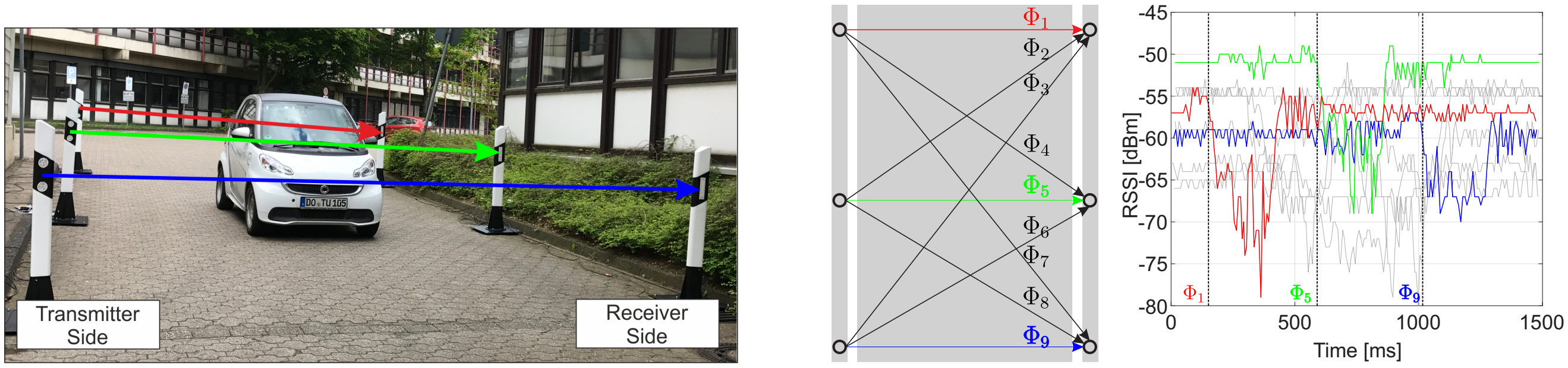

(a) (b)

Fig. 3. Experimental measurement system, schematic and resulting temporal behavior of the different signal attenuations. The direct links are highlighted.

TABLE I
SYSTEM PARAMETERS

| Parameter | Value |
|---|---|
| Transmission frequency | 2.4 GHz |
| Transmission power | 2.5 dBm |
| Antenna type | omnidirectional |
| Antenna height | 1 m |
| Distance between transmitter and receiver | 7 m |
| Longitudinal post distance | 5 m |

### B. Ray-Tracing-based Evaluation and Optimization

Prior to the field measurements, we evaluated the most suitable system settings like orientation, height or type (directional, omnidirectional) of the antennas beeing used as transmitters and receivers with the help of extensive ray tracing simulations. Fig. 4 illustrates a snapshot of an example simulation scenario in which a vehicle passes the proposed system and shadows the Line of Sight (LOS) path between transmitter *TX 2* and the receiving units *RX 1*, *RX 2* and *RX 3*. Consequently, the RSSI values on the receiver side vary depending on which part of the vehicle causes the signal shadowing.

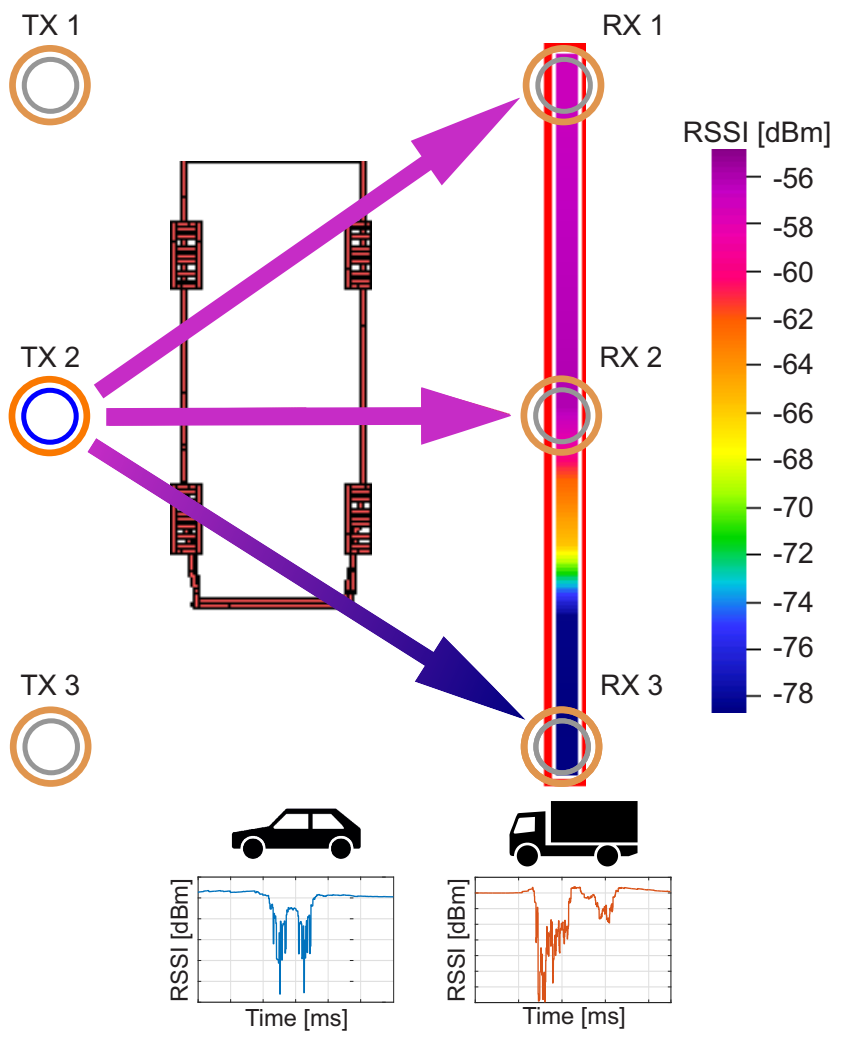

Fig. 4. Example ray tracing scenario illustrating the shadowing effects caused by a vehicle passing the proposed system.

The impact of different antenna heights on the RSSI level when a vehicle is passing the proposed system is illustrated in Fig. 5. For this simulative analysis, the antenna heights of transmitters and receivers are identical. Apparently, there is a correlation between the antenna height and the resulting RSSI level spread. Hence, larger antenna heights lead to an increase of the RSSI level spreads. This is due to the very different vehicle shapes, including the distance between the vehicle's body and the surface of the road.

### C. Feature Extraction and Vehicle Classification at the Master Gateway

One of the receiver posts works as master gateway, aggregating signal information and performing the online-classification process using machine learning. Subsequently, the results of the classification are transmitted to cloud-based services via a mobile radio interface. The master gateway is also responsible for the preprocessing steps that extract the features used for the different machine learning models from the raw signal data. With $d(\Phi_i, \Phi_j)$ being the longitudinal distance between the respective receiver nodes and $\Delta t(\Phi_i, \Phi_j)$ being the time difference of signal drops for two signals $\Phi_i$ and $\Phi_j$, the mean vehicle velocity $\tilde{v}$ is estimated with Eq. 1.

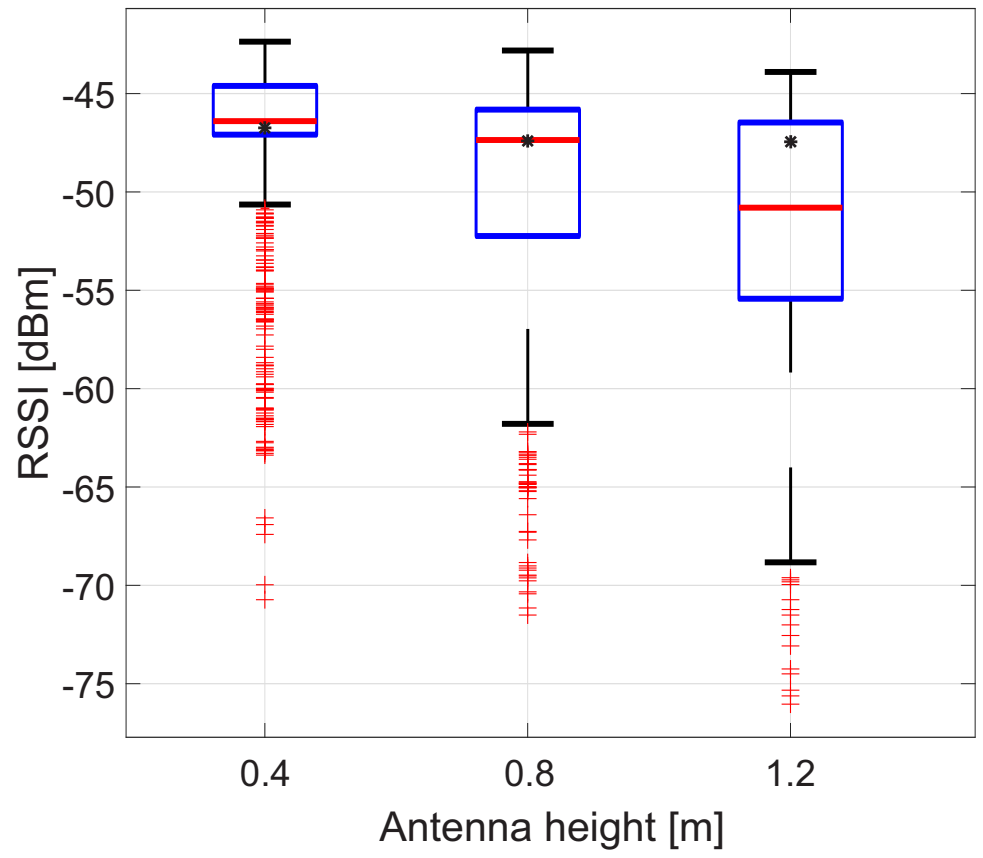

Fig. 5. Impact of different antenna heights on the received signal strength for different vehicle types passing the proposed system evaluated by ray-tracing simulation.

$$\tilde{v} = \frac{1}{3}\left(\frac{d(\Phi_1,\Phi_5)}{\Delta t(\Phi_1,\Phi_5)} + \frac{d(\Phi_1,\Phi_9)}{\Delta t(\Phi_1,\Phi_9)} + \frac{d(\Phi_5,\Phi_9)}{\Delta t(\Phi_5,\Phi_9)}\right) \quad (1)$$

As further features, the duration $t_{drop}$ of the attenuation phase as well as the bulge $b$ of the signal $\Phi$ are computed.

$$b = \frac{1}{N}\sum_{i=1}^{N}\left(\frac{\Phi(i)-\bar{\Phi}}{std(\Phi)}\right)^4 \quad (2)$$

The computed value of $\tilde{v}$ is then used to estimate the vehicle length $\tilde{l}$ using Eq. 3. Both calculations assume constant velocity values of the vehicle passing the system and are therefore affected by an estimation error $\epsilon$.

$$\tilde{l} = \frac{\tilde{v}}{3}\left(t_{drop}(\Phi_1) + t_{drop}(\Phi_5) + t_{drop}(\Phi_9)\right) \quad (3)$$

The magnitude $m$ of the attenuation phase is defined as the difference between the idle level of the signal $\Phi_{idle}$ and the absolute measured minimum. Additionally, the local magnitude $m_l$ is computed as the difference of the maximum and minimum values during the attenuation phase. Furthermore, all local minimum values with a magnitude below the threshold $m_{80} = m * 0.8$ are counted in the variable $n$.

$$\vec{F} = \left\{\tilde{v}, \tilde{l}, t_{drop}, b, m, m_l, n\right\} \quad (4)$$

The resulting feature vector $\vec{F}$ containing the six features is than assembled as shown in Eq. 4. Fig. 6 illustrates the signal-based features on an example plot. For the actual binary classification between the two classes *car* and *truck*, the regression models Artificial Neural Network (ANN), k-Nearest Neighbor (k-NN), Support Vector Machine (SVM) and Decision Tree (DT) are used.

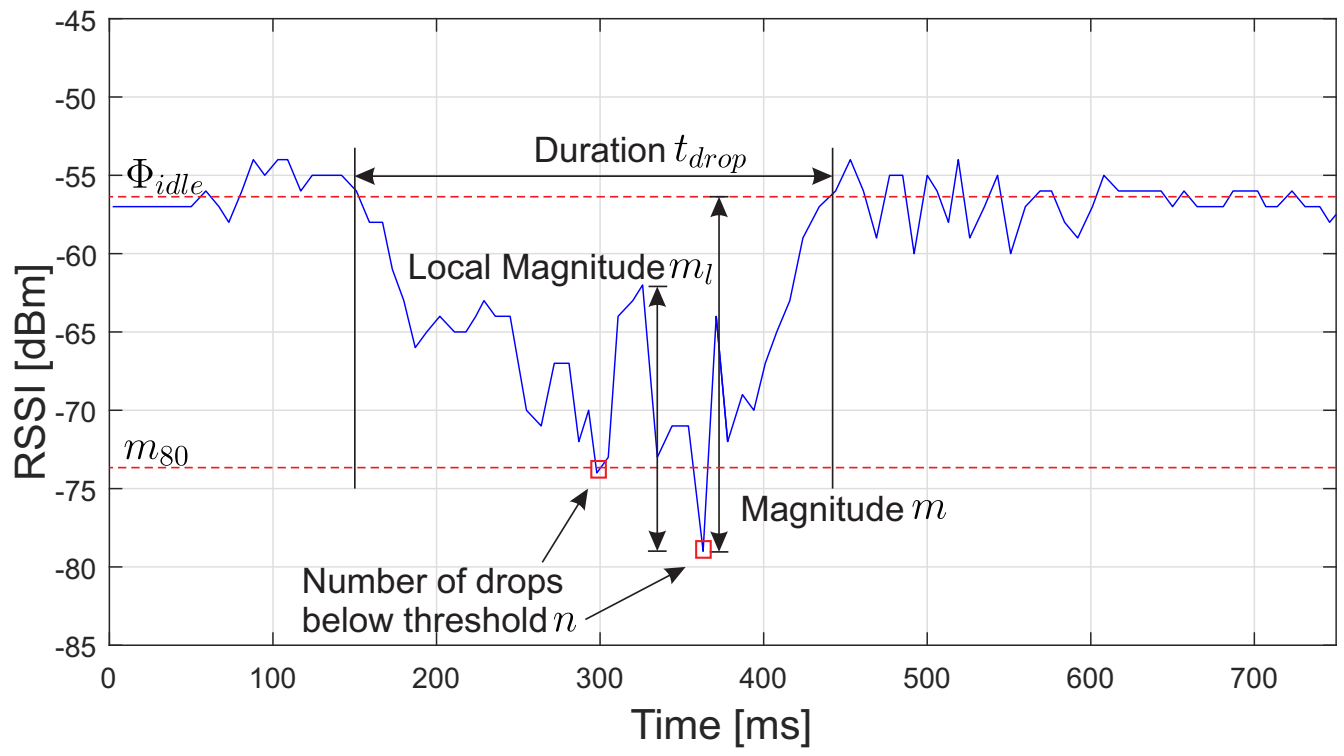

Fig. 6. Visualization of the features for the machine learning process extracted from the RSSI signal. The curve shows a measured example trace for a car passing a single radio link $\Phi$.

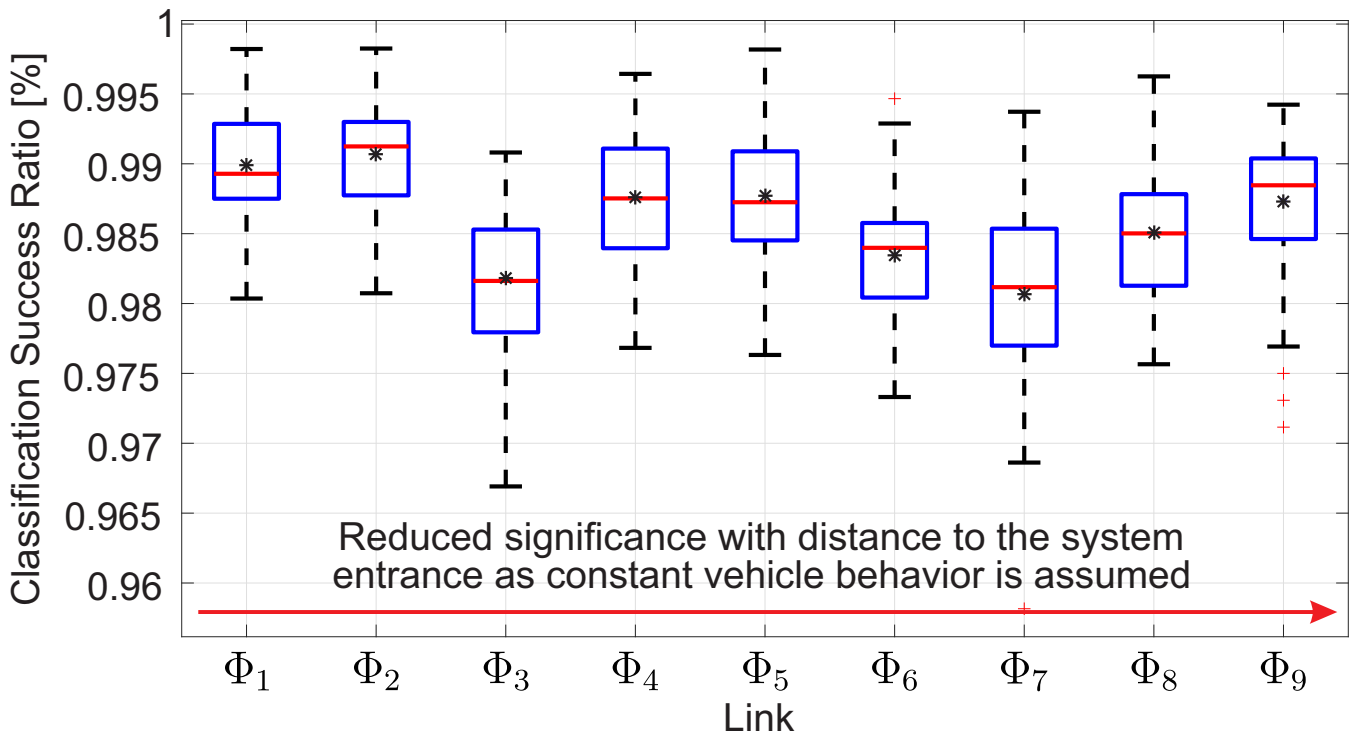

Fig. 7. Classification success ratio for the individual links using k-NN. The direct links are $\Phi_1$, $\Phi_5$ and $\Phi_9$.

## III. First Results of the Empirical Performance Evaluation

In this section, we present first results achieved with our experimental setup. For the following evaluations, we analyzed data from more than 3000 vehicles which was measured on an experimental live deployment of the proposed system at the entrance of a rest area on the German highway A9 within an official test field by the German Federal Ministry of Transport and Digital Infrastructure.

### A. Vehicle Classification

Fig. 7 shows the classification success ratio of the k-NN model for the individual links $\Phi_i$. As the system assumes constant vehicle movements from the moment the vehicle passes the first delineator post, the Classification Success Ratio (CSR) is slightly reduced with increased longitudinal post distance values. Nevertheless, the direct links $\Phi_1$, $\Phi_5$ and $\Phi_9$ can be considered as the most significant ones, as they achieve local maximum values of the CSR with respect to the longitudinal distance. The resulting classification success ratios for the different machine learning models for $\Phi_1$ are shown in Tab. II, proving the high reliability of the proposed vehicle classification system. The highest absolute CSR was reached with the k-NN approach for raw data with 99.07%. However, the classification error ratio of about 1% is caused by vehicles whose shapes cannot be clearly distinguished from those of cars and trucks (e.g. vans). Therefore, the error ratio has to be further reduced if the proposed system should also be used for a large-scale multi-type vehicle classification.

TABLE II
Classification success ratio using 5-fold cross-validation for $\Phi_1$ with regression models: k-Nearest Neighbor (k-NN), Support Vector Machine (SVM), Artificial Neural Network (ANN), Support Vector Machine (SVM). Best values are highlighted.

| | **DT** | **k-NN** | **SVM** | **NN** |
|---|---|---|---|---|
| Raw Data | 97.87% | **99.07%** | 95.91% | 98.59% |
| Feature Vector | 98.61% | 98.99% | 96.03% | **99.05%** |

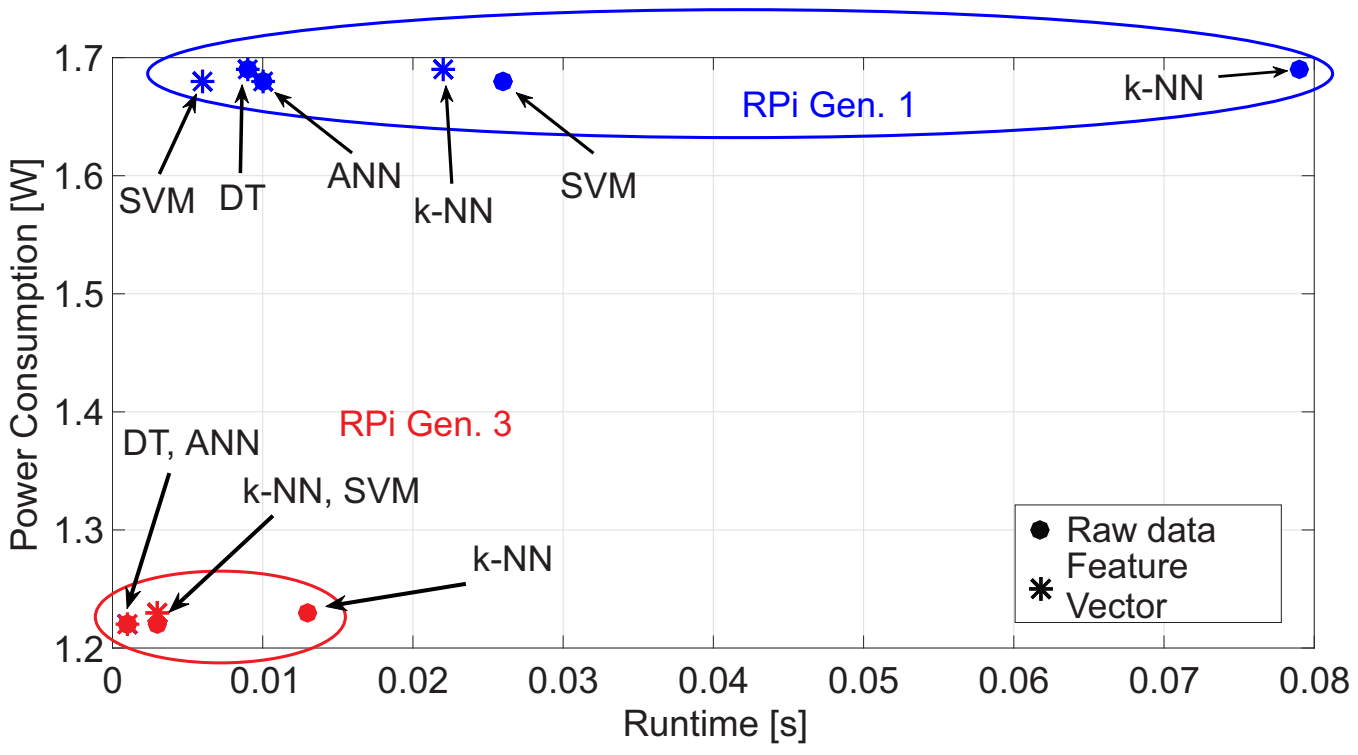

Fig. 8. Energy consumption and computation runtime for online classification with different machine learning models.

### B. Computation Times and Energy-measurements

In order to evaluate the capability of the classification system to be run on small devices, the power consumption was measured with a PowerScale module for Raspberry Pi (RPi) Gen. 1 and Gen. 3 embedded computers. Fig. 8 shows the energy consumption and computation time for the considered machine learning models on the different devices. Obviously, the absolute power consumption is much lower for the RPi Gen. 3 device. For the actual learning models, the consumed energy is on a similar level with k-NN having the highest energy-demand and the highest computation time. Furthermore, it strongly depends on the machine learning model if the feature vector should be used for the classification or the raw signal values. Only the k-NN shows a significantly reduced computation time for the feature vectors. For the other models, the raw data can be directly used in order to avoid further preprocessing steps without negative implications on the consumed resources. The obtained insights in the system provide the way for the next step towards energy autarky which is the migration from embedded computer systems to a fully integrated microcontroller solution. From a resource-efficiency point of view, the best performance can be achieved with the feature vector based ANN, which is also able to achieve a similar high CSR as the k-NN approach.

## IV. Conclusion

In this paper, we presented the system model of a vehicle classification system based on radio-fingerprinting serving as a real-time traffic data source for smart city ITS applications. In contrast to existing approaches, such as camera-based systems, the proposed radio-based approach provides high accuracy, low installation and maintenance costs, weather independence as well as privacy-preservation. The classification performance was evaluated in comprehensive field measurements using an experimental deployment of the system. For the binary classification task of cars and trucks we were able to achieve a classification success ratio of above 99%. From the evaluations, the ANN model provides the best tradeoff between resource-efficiency, real-time capability and CSR. In future work, we will further improve the classification accuracy using a sensor fusion approach to enable a reliable large-scale application of a multi-type vehicle classification system. Furthermore, we will migrate from an embedded computer system integration to a pure microcontroller-based solution approach and focus on optimizing the resource efficiency of the proposed system.

## Acknowledgment

Part of the work on this paper has been supported by the German Federal Ministry for Economic Affairs and Energy as part of the cooperation project between Wilhelm Schröder GmbH, TU Dortmund and FH Dortmund (grant agreement number ZF4038101DB5), and by Deutsche Forschungsgemeinschaft (DFG) within the Collaborative Research Center SFB 876 “Providing Information by Resource-Constrained Analysis”, project B4 “Analysis and Communication for Dynamic Traffic Prognosis”.